\documentclass{article}

\usepackage{amsmath}
\usepackage{graphicx}
\usepackage{graphics}
\usepackage{amssymb}
\usepackage[a4paper]{geometry}
\newtheorem{theorem}{Theorem}[section]
\newtheorem{lemma}[theorem]{Lemma}

\begin{document}

\title{SEMI-ANALYTIC EQUATIONS TO THE COX-THOMPSON INVERSE SCATTERING METHOD AT
FIXED ENERGY FOR SPECIAL CASES}

\author{TAM\'AS P\'ALMAI\\
Department of Theoretical Physics, Budapest University of Technology and Economics\\
H-1111, Budapest,
Hungary\\
palmai.tamas@gmail.com\\ \ \\
MIKL\'OS HORV\'ATH\\
Institute of Mathematics, Budapest University of Technology and Economics\\
H-1111, Budapest,
Hungary\\
horvath@math.bme.hu\\ \ \\
BARNAB\'AS APAGYI\\
Department of Theoretical Physics, Budapest University of Technology and Economics\\
H-1111, Budapest,
Hungary\\
apagyi@phy.bme.hu}

\maketitle

\begin{abstract}
Solution of the Cox-Thompson inverse scattering problem at fixed energy\cite{Cox1970,Apagyi2003,Melchert2006} is reformulated resulting  in semi-analytic equations. The new set of equations for the normalization constants and the nonphysical (shifted) angular momenta are free of matrix inversion operations. This simplification is a result of treating only the input phase shifts of partial waves of a given parity. Therefore, the proposed method can be applied for identical particle scattering of the bosonic type (or for certain cases of identical fermionic scattering). The new formulae are expected to be numerically more efficient than the previous ones. Based on the semi-analytic equations an approximate method is proposed for the generic inverse scattering problem, when partial waves of arbitrary parity are considered.
\end{abstract}

\section{Introduction}
Inverse quantum scattering theories represent an  important field of physics research for more than fifty years.
The reason for this ever growing interest is simple: application of the inverse methods may yield information about the interactions governing the microscopic processes which can be detected by studying simple macroscopic scattering experiments.

There are several theories of inverse scattering; in this paper we treat only one of them, namely the Cox-Thompson (CT) method at fixed energy\cite{Cox1970}. This procedure produces potentials that possess finite values at the origin and non-zero first momenta. Furthermore the CT scheme for inverse scattering at fixed energy proves to be a very efficient constructive procedure ever proposed, although it becomes difficult to carry it through if input data possess large errors. Thus simplifications of this method is a question of interest.

In this paper we show how the CT method can be simplified when it is applied to identical particle scattering  of bosonic type. In this case only the even partial wave ($\ell=0,2,...$) phase shifts can be extracted from the experimental data and the solution of the nonlinear CT equations can be simplified. The simplification means here that while the general nonlinear system of the CT equations contains implicit matrix inversions, the simplified equations do not. In the case of even partial waves (and also for the odd waves, $\ell=1,3,...$) the matrix inversions can be performed analytically. Therefore the simplified method is called as 'semi-analytical' one.

Explicit calculations show that the new method provides more stable solutions to the nonlinear CT equations. Also, it turns out that the separate treatment of odd and even partial waves and the subsequent addition of the resulted two potentials is a good approximation.

In the next section we review the CT method. The simplified equations valid in case of even or odd angular momenta are derived in section 3 and various applications are presented in section 4. Section 5 is left for a short summary.

\section{Cox-Thompson method}

We consider the   scattering of particles by a spherically symmetric potential. The main equation of our interest is thus the radial Schr\"odinger equation. The well-known partial wave decomposition leads to
\begin{equation}
\left(-\frac{\hbar^2}{2m}\frac{1}{r}\frac{d^2}{dr^2}r+\frac{\hbar^2}{2m}\frac{l(l+1)}{r^2}+V(r)-E\right)\frac{\psi_l(r)}{r}=0,\qquad l\in S
\end{equation}
for the $\psi_l(r)$ partial waves involved in the scattering event.

Let us introduce the commonly used dimensionless quantities, $x=kr$ for the distance, and $q(x)=\frac{V(x/k)}{E}$ for the potential where $E= {\hbar^2k^2}/{2m}$ denotes the scattering energy, $m$ the reduced mass, and $k$ the wave number.

One can rewrite the radial Schr\"odinger equation into a more appealing form
\begin{equation}\label{DSE}
D_q(x)\psi_l(x)=l(l+1)\psi_l(x),\qquad l\in S,
\end{equation}
where the differential operator  $D_q(x)$ is defined as
\begin{equation}
D_q(x)=x^2\left(\frac{d^2}{dx^2}+1-q(x)\right).
\end{equation}
Clearly in case of a zero potential the two linearly independent solutions of the Schr\"odinger equation  (\ref{DSE}) are the Riccati-Bessel and the Weber-Schl\"afli functions defined as
\begin{equation}\label{defRB}
u_l(x)=\sqrt{\frac{\pi x}{2}}J_{l+\frac{1}{2}}(x),\qquad v_l(x)=\sqrt{\frac{\pi x}{2}}Y_{l+\frac{1}{2}}(x).
\end{equation}
It is a well-known fact that one can transform the free solutions $u_l(x)$ into the solutions $\psi_l(x)$ by using  a generalized translation operation\cite{Chadan}:
\begin{equation}\label{PL}
\psi_l(x)=u_l(x)-\int_{0}^{x}dtt^{-2}K(x,t)u_l(t), \qquad l\in S.
\end{equation}
This equation is often referred to as the Povzner-Levitan representation of the wave function, the quantity $K(x,y)$ is named as the transformation kernel.

The inverse potential is calculated from the transformation kernel as follows
\begin{equation}\label{pot}
q(x)=-\frac{2}{x}\frac{d}{d x}\frac{K(x,x)}{x}.
\end{equation}
The transformation kernel is determined by an integral equation of
 the Gel'fand-Levitan--type (GL--type)   which contains the symmetric input kernel $g(x,y)$\footnote{The requirements for the introduced input and output kernel quantities can be found in Ref\cite{Chadan}. We note that the Cox-Thompson expansions meet these criteria.}:
\begin{equation}\label{GL}
K(x,y)=g(x,y)-\int_{0}^{x}d tt^{-2}K(x,t)g(t,y),\qquad x\geq y.
\end{equation}
In the CT method we take the following separable expansion for the input kernel:
\begin{equation}\label{CTang}
g(x,y)=\sum_{l\in S}\gamma_lu_l(x_<)v_l(x_>), \qquad x_{\left\{\begin{subarray}{l}<\\>\end{subarray}\right\}} = \left\{ \begin{subarray}{l}\min\\ \max\end{subarray} \right\} (x,y)
\end{equation}
where $\gamma_l$'s are expansion coefficients (not required to be determined), $u_l$'s and $v_l$'s are defined by Eq. (\ref{defRB}). The summation set $S$ consists of the physical angular momenta $\{l\}$.

The requirements for the input kernel are i) to satisfy a differential equation (with a specific boundary condition)\cite{Chadan} and ii)  to be uniquely solvable by the GL-type integral equation (\ref{GL}) at hand. While the first condition can be easily checked the second one is proved  only for continuous transformation kernels\cite{Coxuniq}. But this is not a restriction since $K(x,y)$ is a twice continuously differentiable function.

To solve the GL equation (\ref{GL}), Cox and Thompson\cite{Cox1970} introduced a separable
ansatz for the transformation kernel, in the form of a sum over an artificial angular momentum space $L\in T$
\begin{equation}\label{CTanK}
K(x,y)=\sum_{L \in T}A_L(x)u_L(y)
\end{equation}
where $A_L$'s are unknown expansion functions and  $T$ is interpreted as a set of unknown shifted angular
momenta to be determined under the constraint that it has the same number $N$ of  {\it different} elements
as $S$ does, i.e., $N=|S|=|T|$ and $S\cap T=\emptyset$.

By inserting the ans\"atze (\ref{CTang}) and (\ref{CTanK}) into the GL equation (\ref{GL}) and using the linear
independence of the Riccati-Bessel functions, one obtains equations for the determination of the
expansion functions $A_L(x)$ as follows
\begin{equation}\label{AL}
\sum_{L\in T}A_L(x)\frac{W[u_L(x),v_l(x)]}{l(l+1)-L(L+1)}=v_l(x)\qquad l\in S.
\end{equation}
Here, the only unknowns are the set $T$ with elements $L$, and $W$ denotes the Wronskian defined by $W[a,b]\equiv ab'-a'b$.

It can also be proved that Eq. (\ref{AL}) is uniquely solvable for continuous expansion functions, therefore the ansatz for the transformation kernel (\ref{CTanK}) does not restrict the solution of the GL-type integral equation.

If the set $T$  is given one only has to solve a system of linear equations and perform a derivation in order to create the inverse potential (\ref{pot}). Consequently the non-trivial task is to obtain $T$.

In order to determine the set $T$ one makes use of the Povzner-Levitan representation of the radial
scattering wave function (\ref{PL}) whose asymptotic form, containing the input phase shift data, takes the form
\begin{equation}\label{PLa}
B_l\sin(x-l\frac{\pi}{2}+\delta_l)=\sin(x-l\frac{\pi}{2})-\sum_{L\in
T}A_L^{\rm{a}}(x)\frac{\sin((l-L)\frac{\pi}{2})}{l(l+1)-L(L+1)},\quad
l\in S.
\end{equation}
Here, $B_l$'s are normalization constants and we have defined the
asymptotic expansion functions $A_L^{\rm{a}}(x)\equiv
A_L(x\to\infty)$ which can be calculated from the asymptotic
version of Eq. (\ref{AL}) which is given by
\begin{equation}\label{ALa}
\sum_{L\in T}A_L^{\rm{a}}(x)\frac{\cos((l-L)\frac{\pi}{2})}{l(l+1)-L(L+1)}=
-\cos(x-l\frac{\pi}{2}),\qquad l\in S.
\end{equation}
Using the last two equations, (\ref{PLa}) and (\ref{ALa}), one can easily derive  the following equations for the determination of the unknown $L$'s from the input phase shifts, $\delta_l$'s\cite{Apagyi2003,Melchert2006}:
\begin{equation}\label{general}
S_l=\frac{1+i\mathcal{K}_l^+}{1-i\mathcal{K}_l^-}\qquad {\rm{or}}
\qquad\ \tan(\delta_l)=\frac{\mathcal{K}_l^++\mathcal{K}_l^-}{2+i(\mathcal{K}_l^+-\mathcal{K}_l^-)},
\qquad l\in S
\end{equation}
where $S_l=e^{2i\delta_l}$ and the "shifted" reactance matrix elements are defined as
\begin{equation}
\mathcal{K}_l^{\pm}=\sum_{L\in T, l'\in S}[M_{\sin}]_{lL}[M_{\cos}^{-1}]_{Ll'}e^{\pm i(l-l')\pi/2},
\qquad l\in S,
\end{equation}
with
\begin{equation}\label{Msincos}
\left\{\begin{array}{ll}
M_{\sin}\\
M_{\cos}
\end{array}\right\}_{lL} = \frac{1}{L(L+1)-l(l+1)}\left\{ \begin{array}{ll}
\sin\left((l-L)\frac{\pi}{2}\right)\\
\cos\left((l-L)\frac{\pi}{2}\right)
\end{array} \right\},\qquad l\in S,\,L\in T.
\end{equation}

Clearly, the appearance of inverse of matrix $M_{\cos}$ makes solutions of Eq. (\ref{general}) for the unknowns $L$ especially hard. In practice this means that finding its solution is not an easy task for any nonlinear solvers.

But, for instance with the deployment of sophisticated numerical methods, that can be applied to such ill-conditioned scenarios, one may succeed to solve either one or both\footnote{Despite the two sets of equations are equivalent with each other, for practical scenarios they are very different regarding their degree of ill-condition.} of the highly nonlinear equations (\ref{general}) and thus find the set $T$. From that, as mentioned before, the CT inverse potential $q (x)$ can easily be obtained by employing Eqs. (\ref{AL}), (\ref{CTanK}), and (\ref{pot}).

\section{Semi-analytic solution}

In this section we present simplifications to Eqs. (\ref{general}) which can be used if only
even (odd) partial waves are arising during the collision. The simplified equations can
be employed also to construct a fair approximation to treat generic scattering problems. These approximations
 will be discussed in a later section.

By solving Eq. (\ref{ALa}) explicitly without matrix inversion we will arrive to a simple set of nonlinear equations for the set $T$.
First we will show that Eq. (\ref{ALa}) can be solved uniquely. This uniqueness is a consequence of the following statement.

\begin{lemma}
For $x_{i},y_{j}\in\mathbb{C}\quad x_i\neq y_j\quad \forall i,j\in I$ ($|I|=N$) the system
\begin{equation}\label{unique}
\sum_{i\in I}\frac{a_{i}}{y_{j}-x_{i}}=0,\qquad j\in I
\end{equation}
has the unique solution $a_{i}\equiv 0\quad \forall i\in I$.
\end{lemma}

In the proof of the uniqueness we consider two solution sets ($\{A_L^{\rm a}(x)\}$ and $\{B_L^{\rm a}(x)\}$) which satisfy Eq. (\ref{ALa}). For the differences $\{A_L^{\rm a}(x)-B_L^{\rm a}(x)\}$ we have a set of equations of the type (\ref{unique}). Hence because of Lemma 1 the difference between two sets of solutions is zero which means the uniqueness of the solution of Eq. (\ref{ALa}).

Now let us differentiate Eq. (\ref{ALa}) twice with respect to the variable $x$. Then by employing the uniqueness result we arrive at the following equation
\begin{equation}\label{difA}
\frac{d^2A_L^{\rm{a}}(x)}{d x^2}=-A_L^{\rm{a}}(x),
\end{equation}
which has a periodic solution as
\begin{equation}\label{difAsol}
A_L^{\rm{a}}(x)=a_L\cos(x)+b_L\sin(x).
\end{equation}
Now, by inserting this solution (\ref{difAsol}) into Eq. (\ref{ALa}) and taking into account
the independence of the sine and cosine functions, one gets the following two sets of equations
\begin{equation}\label{albl}
\sum_{L\in T}\left\{\begin{array}{ll}
a_L\\
b_L
\end{array}\right\}\frac{\cos\left((l-L)\frac{\pi}{2}\right)}{L(L+1)-l(l+1)} = \left\{\begin{array}{ll}
\cos\left(l\frac{\pi}{2}\right)\\
\sin\left(l\frac{\pi}{2}\right)
\end{array} \right\},\qquad l\in S.
\end{equation}

Consider two distinct cases of input data. In the first case let $S$ consist of only even numbers (henceforth let us denote such a set with $S_{\rm e}$) and in the second only of odd numbers ($S_{\rm o}$). In these  two distinct cases, instead of (\ref{albl}), we have two kinds of equations (notice the simplification in comparison to (\ref{albl})):
\begin{equation}\label{evene}
\sum_{L\in T_{\rm{e}}}\left\{\begin{array}{ll}
a_L\\
b_L
\end{array}\right\}\frac{\cos\left(L\frac{\pi}{2}\right)}{L(L+1)-l(l+1)} = \left\{\begin{array}{ll}
1\\
0
\end{array} \right\},\qquad l\in S_{\rm{e}}
\end{equation}
and
\begin{equation}\label{odde}
\sum_{L\in T_{\rm{o}}}\left\{\begin{array}{ll}
a_L\\
b_L
\end{array}\right\}\frac{\sin\left(L\frac{\pi}{2}\right)}{L(L+1)-l(l+1)} = \left\{\begin{array}{ll}
0\\
1
\end{array} \right\},\qquad l\in S_{\rm{o}},
\end{equation}
where $|T_{\rm{e}}|=|S_{\rm{e}}|$, $|T_{\rm{o}}|=|S_{\rm{o}}|$ and
$T_{\rm{e}}\cap S_{\rm{e}}=\emptyset$, $T_{\rm{o}}\cap
S_{\rm{o}}=\emptyset$. In order to solve these sets of equations we need another statement similar to the one for the sets of equations of type (\ref{unique}).

\begin{lemma}
For $x_{i},y_{j}\in\mathbb{C}\quad x_i\neq y_j\quad \forall i,j\in I$ ($|I|=N$) the system
\begin{equation}
\sum_{i\in I}\frac{a_{i}}{y_{j}-x_{i}}=-1,\qquad j\in I
\end{equation}
has the unique solution
\begin{equation}
a_{k}=\frac{\prod_{i\in I}(x_{k}-y_{i})}{\prod_{i\in I\setminus \{k\}}(x_{k}-x_{i})},\qquad k\in I.
\end{equation}
\end{lemma}
Employing Lemma 1 and 2 for Eqs. (\ref{evene}) and (\ref{odde}) one can arrive at the solutions
\begin{equation}\label{al}
a_L=\frac{\prod_{l\in S_{\rm{e}}}(L(L+1)-l(l+1))}{\prod_{L'\in T_{\rm{e}}\backslash\{L\}}
(L(L+1)-L'(L'+1))}\frac{1}{\cos\left(L\frac{\pi}{2}\right)},\qquad b_L=0,\qquad L\in T_{\rm{e}},
\end{equation}
and
\begin{equation}\label{bl}
a_L=0,\qquad b_L=\frac{\prod_{l\in S_{\rm{o}}}(L(L+1)-l(l+1))}{\prod_{L'\in T_{\rm{o}}\backslash\{L\}}
(L(L+1)-L'(L'+1))}\frac{1}{\sin\left(L\frac{\pi}{2}\right)},\qquad L\in T_{\rm{o}},
\end{equation}
respectively.

Now, by using the explicit expressions (\ref{al}) in Eqs. (\ref{difAsol})  and (\ref{PLa}), one obtains finally the simplyfied 'semi-analitic' equations to the CT method as
\begin{equation}\label{even}
\tan(\delta_{l})=-\sum_{L \in T_{\rm{e}}}\frac{\prod_{l'\in
S_{\rm{e}}\backslash\{l\}}(L(L+1)-l'(l'+1))}{\prod_{L'\in
T_{\rm{e}}\backslash\{L\}}(L(L+1)-L'(L'+1))} \tan\left(L\frac{\pi}{2}\right),\qquad
l\in S_{\rm{e}},
\end{equation}
valid for the case of even $l$'s. Similarly, using Eqs. (\ref{bl}) we get the semi-analytic equations to the CT method as
\begin{equation}\label{odd}
\tan(\delta_{l})=\sum_{L \in T_{\rm{o}}}\frac{\prod_{l'\in
S_{\rm{o}}\backslash\{l\}}(L(L+1)-l'(l'+1))}{\prod_{L'\in
T_{\rm{o}}\backslash\{L\}}(L(L+1)-L'(L'+1))} \cot\left(L\frac{\pi}{2}\right),\qquad
l\in S_{\rm{o}}
\end{equation}
which are valid in the case of odd $l$'s. These equations determine the unknown set $T_{\rm{e}}$
or $T_{\rm{o}}$ of shifted angular momenta $L$.\footnote{One can also determine the normalization constants. In both cases the result is $B_l=[\cos(\delta_l)]^{-1}$.}

Notice the simplified structure of the nonlinear equations (\ref{even}) and (\ref{odd}), compared to Eqs. (\ref{general}). While Eqs. (\ref{general}) contain an explicit matrix inversion of a matrix involving the unknowns of shifted angular momenta, $L$'s, formulae (\ref{even}) and (\ref{odd}) do not require such a nonlinear operation (hence we use the term semi-analytic equations for it). They 'only' contain products and the tangent (cotangent) operations and are thus presumably easier to be solved for the sets $T_{\rm{e}}$ or $T_{\rm{o}}$, if the respective input phase shifts are given.

Finding the sets $T_{\rm{e}}$ or $T_{\rm{o}}$, the corresponding potentials $q_{\rm{e}}(x)$ or $q_{\rm{o}}(x)$ can be obtained similarly as in the general case, by employing Eqs. (\ref{AL}), (\ref{CTanK}), and (\ref{pot}).

\section{Applications}

As mentioned in the introduction the method discussed in section 3 can be applied for identical particle scattering of bosonic and fermionic type. It was checked that with the new equations such problems can be handled easily and without the deployment of sophisticated numerical methods, like simulated annealing\cite{Schumayer2008}.

Here we introduce another application of our method. Regarding the generic inverse scattering problem consider the decomposition
$$
S=S_{\rm{e}}\cup S_{\rm{o}}
$$
where $S_{\rm{e}}$ and $S_{\rm{o}}$ contains, respectively, the even and odd elements of $S$. For the sets $S_{\rm{e}}$ and $S_{\rm{o}}$ one can perform the semi-analytic process. The resulting potentials are $q_{\rm{e}}(x)$ and $q_{\rm{o}}(x)$, respectively.

Let us treat the general CT   equations (\ref{general}) as a set of coupled equations for the elements of $T$. Now if one disregards the coupling between the equations indexed by even and odd physical angular momenta then one gets an approximate method. It is obvious that in this approximation $T_{\rm{e}}\cup T_{\rm{o}}$ is obtained for the set of the shifted angular momenta but let us use the  sets $T_{\rm{e}}$ and $T_{\rm{o}}$ separately in the following.

Considering, moreover, the simple additional prescription in equation (\ref{CTanK}), we conclude that it is reasonable to approximate the potential $q(x)$ as
$$
q_{\rm a}(x)=q_{\rm{e}}(x)+q_{\rm{o}}(x).
$$

Since the solution of the semi-analytic equations is expected easier to be solved than the ones arising in the original procedure, the proposed approximation could be used for a larger class of input data. Below we demonstrate the applicability of the proposed approximation through inversion of experimental phase shift data.

Chen and Thornow\cite{Chen2005} have performed a comprehensive study of $n$ scattering by $^{12}$C target nucleus in the  energy region 7 MeV $\le E_{\rm{n}}^{\rm{lab}}\le$ 24 MeV. They have derived 88 sets of complex-valued phase shifts, one of which we use in our calculation here. Because of the spin-orbit coupling, each partial wave provides two phase shifts, $\delta_l^+$ and $\delta_l^-$. In case of a weak spin-orbit coupling the combined phase shifts $\delta_l=[(l+1)\delta_l^+ +l\delta_l^-]/(2l+1)$ are characteristic of the underlying central potential\cite{Leeb1995}.

Our input data will be the set of phase shifts at energy $E_{\rm{n}}^{\rm{lab}}=12$ MeV. In Table 1 we listed the quantities $\delta_l^{\rm{orig}}={\rm Re}{\delta_l}$, the real parts of the combined phase shifts, and also $\eta_l^{\rm{orig}}=|\exp(2i\delta_l)|$, the elasticities. The results of the general CT procedure and the approximate method are visualized in Fig. 1.

\begin{table}[h]
\caption{Original phase shifts $\delta_l^{\rm{orig}}$ and elasticities $\eta_l^{\rm{orig}}$ taken from the experimental phase shift data [see the text] for  ${n}$ scattering by $^{12}$C nucleus at the energy $E_{\rm{n}}^{\rm{lab}}=12$ MeV. Both the real and imaginary parts of the shifted angular momenta $L$ corresponding to the general (g) and the approximate (a) method, and the original phase shifts with the absolute differences between them and the calculated ones are listed.}

\begin{center}
\begin{tabular}{@{}ccccccccccc@{}}
\hline
$l$  & ${\rm Re}L_{\rm g}$&${\rm Im}L_{\rm g}$&${\rm Re}L_{\rm a}$&${\rm Im}L_{\rm a}$& $\delta_l^{\rm orig}$&$\eta_l^{\rm orig}$& $\Delta_l^{\rm g}$&$\Xi_l^{\rm g}$& $\Delta_l^{\rm a}$&$\Xi_l^{\rm a}$ \\ \hline
0 & -0.615            & -0.068            & -0.516            &\hphantom{-}0.010&\hphantom{-}0.522&0.580&0.042&0.043&0.039&0.125\\
1 & \hphantom{-}1.152 & \hphantom{-}0.011 & \hphantom{-}1.480 &-0.033           &-0.737           &1.000&0.044&0.130&0.303&0.383\\
2 & \hphantom{-}2.613 & -0.338            & \hphantom{-}2.476 &-0.209           &-0.689           &0.560&0.130&0.013&0.251&0.005\\
3 & \hphantom{-}2.905 & \hphantom{-}0.037 & \hphantom{-}3.011 &-0.129           &\hphantom{-}0.172&0.643&0.128&0.054&0.163&0.048\\
4 & \hphantom{-}4.099 & -0.146            & \hphantom{-}4.095 &-0.145           &\hphantom{-}0.021&0.831&0.093&0.009&0.018&0.007\\ \hline
\end{tabular}
\end{center}

$\Delta_l^{i}=|\delta_l^{i}-\delta_l^{\rm orig}|$ and $\Xi_l^{i}=|\eta_l^{i}-\eta_l^{\rm orig}|$.
\end{table}

\begin{figure}[h!]
  \label{fig1}
  \caption{Inverse potentials $V(r)$ obtained from input phase shifts  $\delta_l^{\rm{orig}}$  and elasticities  $\eta_l^{\rm{orig}}$  [see Table 1]  as a function of the radial distance $r$   at  energy $E_{\rm{n}}^{\rm{lab}}=12$ MeV ($E_{\rm{n}}^{\rm{c.m.}}=11.08$ MeV, $k=0.766$ fm$^{-1}$). Potential components  obtained  by  the generic method and the approximate one are denoted by solid and dashed curves, respectively.\label{nC}}
  \centering
    \includegraphics[width=0.8\textwidth]{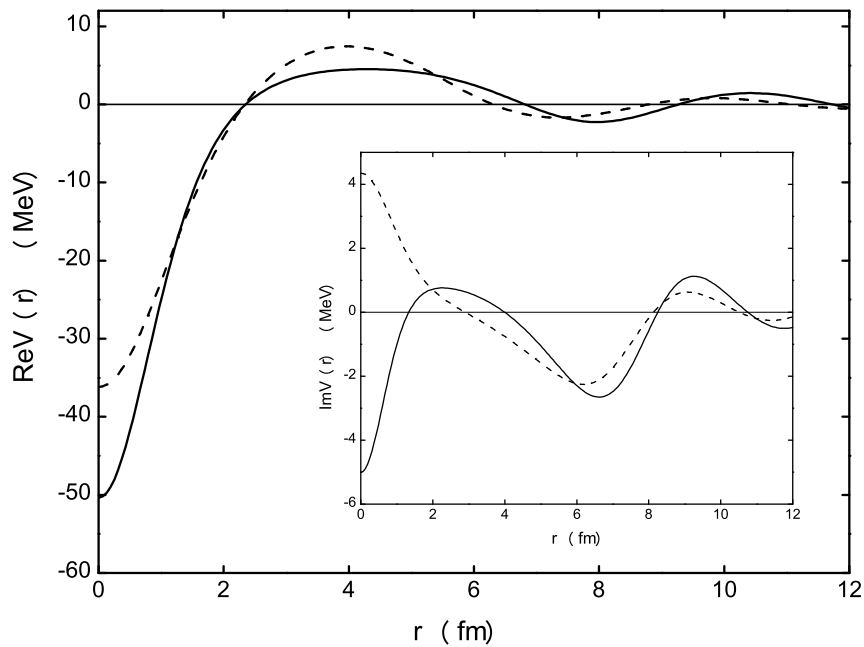}
\end{figure}

To study the applicability of the approximate procedure one can compare the curves corresponding to the different procedures as shown in Fig. 1. We assume that the general method gives the correct result. As it can be seen the results obtained by employing the semi-analytic equations give a fair approximation. Indeed, if one calculates back the phase shifts and elasticities from the inverse potentials $V_{\rm{g}}$ and $V_{\rm{a}}$, we see that the reproduction is the best for the CT potential. Such observations can be made also by studying Table 1 where we have listed the differences $\Delta_l$ and $\Xi_l$ between the original phase shifts $\delta_l^{\rm{orig}}$ and $\eta_l^{\rm{orig}}$ and the re-calculated ones provided  by the inverse potentials of the two different methods. From this and other examples studied so far we conclude that the proposed approximation to the CT method can be used for a global orientation about the nature of the underlying interaction.

\section{Summary}

A simplified version of the main equation (\ref{general})  of the Cox-Thompson inverse scattering method at fixed energy was introduced. The simplification means that unlike the generic equation the new formulae do not involve the inversion of a matrix which contains the variables to be determined. Consequently the new equations (\ref{even}) and (\ref{odd}) are easier to be solved by usual nonlinear solvers (such as that based, e.g., on the Newton-Raphson procedure\cite{NumericalRecipes}). However this  simplification is only possible in cases when the partial waves arising in the scattering event are of the same parity. Thus the new equations are valid for the scattering of identical bosons or fermions. Because, in practice, there are also general collision experiments, we introduced an approximate procedure which is based on the fact that the potential is expressed as a sum for the shifted angular momenta. By an example taken from nuclear physics we demonstrated the applicability of the approximate method.

We note that a general semi-analytical method which can be applied for all physical partial angular momenta is not yet known. Its development would be a promising field of future study.

\end{document}